# Effect of elasto-plastic compatibility of grains on the void initiation criteria in low carbon steel

Anish Karmakar[1*,] Kaustav Barat[2]

[1]Indian Institute of Technology, Roorkee, Uttrakhand-247667, India
[2]CSIR-National Aerospace Laboratories, Bangalore-560017, Karnataka, India
Corresponding author email id: *anishfmt@iitr.ac.in*

**Abstract:**
The present study evidences the role of ferrite grain size distributions on the occurrence of void initiation in a low carbon steel. Various thermomechanical treatments were done to create ultrafine, bimodal and coarse range of ferrite grain distributions. A two parameter characterization of probable void initiation sites is proposed; elastic modulus difference and difference in Schmid factor of the grains surrounding the void. All microstructures were categorized based on the ability to ease or resist void nucleation. For coarse grains, elastic modulus difference as well as the Schmid factor difference is highest, intermediate for ultrafine and lowest for bimodal microstructure.

Tremendous drives are there from automotive, naval and oil/gas sectors for the development of steels with optimum combination of strength, ductility, toughness and weldability [1]. In recent years, demand for electric vehicles has taken the mandate for lightweight steel to a new pedestal. Mechanical properties can be improvised depending upon the tuning of the microstructural constituents [2,3]. The requirement of strength is undoubtedly very much essential from the viewpoint of fatigue and crash resistance of the component [4]. Alongside the requirement of good ductility, formability/weldability can't be neglected [5,6]. In order to design a better weldable and formable component, fully ferritic microstructure is an essential criterion [2,3,7]. Steels with uniform distribution of fine microalloyed carbides / cementite in the ultrafine grained ferrite matrix is developed for optimum balance of strength and ductility [8,9]. Role of carbide particles in the failure behaviour of ferrite-carbide structures has been discussed elaborately in literature [10–12]. At larger carbide contents, the dispersion rate decreases with a coarse distribution of carbides. These carbides crack in front of the stress concentration front as well as facilitating cracks to propagate via grain boundaries [12]. Nanoscale carbide dispersed NanoHiten and BHT steels have been especially designed for car-chassis [13]. The degeneration of carbides has been studied for a long time considering the pile-up of the dislocations against carbide particles, generating stress concentrations in front of the particles [11,14]. The results in [10] evidenced that the toughness measured in notched specimens is mainly determined by the grain sizes, which define the local fracture stress ($\sigma_f$). Size of carbide particle plays a minor role there. However, on the contrary, in pre-cracked specimens the toughness is sensitive to the carbide sizes, which affect the critical plastic strain ($\varepsilon_{pc}$) for initiating a crack nucleus. The role of ferrite matrix on the phenomena of fracture has been subdued under the presence of carbides. Overall the finer carbides with a moderate volume fraction can control the properties better. The effect of ferrite grain size with a bimodal distribution on the tensile properties of low carbon steels have already been studied by several researchers [15–17]. Damage initiation can be expressed from the elasto-plastic incompatibility created during the straining. The elastic incompatibility prior to straining can be assessed by the elastic modulus difference of microstructural domains getting deformed. Similarly, the plastic incompatibility can be given by the Schmid factor difference of the grains involved in deformation. Present study additionally shows that the steel having carbides less than 20 vol. %, the elastic modulus and Schmid factor of different grains around a void can be used to predict potent damage nucleation sites and based on that a criterion for void initiation can be defined. Previous work depicted the concept of incompatibility modulus to explain the elasto-plastic strain evolution in continua containing dislocation [18]. A finite strain can be decomposed to a compatible and incompatible part and if it is a elasto-plastic strain, this can be decoupled to a plastic and elastic part. Experimental determination of this incompatibility modulus is yet to be attempted, therefore; in the present study this problem has been taken up with the help of two readily derivable quantities from EBSD data, i.e. elastic modulus and Schmid factor difference. As described by Baczmanski' et al [19], the variation of elastic grain stresses is caused by the misfit of the elastic strains in grains in relation to the surrounding polycrystalline aggregate. Now, coming to plastic deformation, in order to predict the same, the contribution of slip on crystallographic planes must be considered. The essential criterion used in the model is the Schmid criterion [Eqn. 1] according to which the set of active slips is selected.

$$\sigma^{slip}_{[uvw]<hkl>} = \tau^{slip}_{[uvw]<hkl>} \tag{1}$$

A 6 mm thick steel strip has been used in this present study with a nominal composition of Fe- 0.10 C-0.33 Si-1.42 Mn-0.01 P-0.003 S-0.035 Al-0.05 Nb-0.05 V-0.007 N (wt. %). The alloy is soaked at 1100 °C for 30 minutes followed by different thermomechanical processing schedules as depicted in Fig. 1a. Heavy deformation in the metastable austenite region (~830 °C - between $Ae_3$ and $Ar_3$) followed by air-cooling produces ultrafine ferrite-carbide structures. Heavy warm deformation (~ 600 °C) followed by rapid intercritical annealing generates bimodal ferrite-carbide structure whereas the air cooled sample after 80 % cold rolling and annealing (600 °C for 8 h) is responsible for the coarse one. The mechanisms behind the formation of the microstructures are given elsewhere [15,16,20–22]. The microstructures of the heat treated samples are prepared following standard metallographic techniques. The micrographs are characterized by scanning electron microscope (Zeiss-Auriga) with an attachment of electron backscattered diffraction (EBSD). For EBSD analysis, the samples are electropolished with a solution of methanol and perchloric acid (80:20 ratios) kept at -40 °C. Grain size of each samples are measured by equal circle diameter (ECD) method described in [23,24] considering at least five secondary electron images of each sample. All these structures are having carbide <25 vol. %. Elastic modulus and Schmid factor of different grains around a void have been calculated from TSL-OIM software after EBSD scanning. The standard tensile samples have been machined following ASTM E-8 standard [25] and tested uniaxially in Instron-8862 Universal Testing Machine (10 ton capacity). The band contrast images generated after EBSD scanning of the different ferrite-carbide structures are shown in Fig. 1(b-d). Inset in Fig. 1(c) is showing the bimodal distribution of ferrite grain sizes. Average grain sizes for ultrafine and coarse grain regions are 2.2±0.5 and 10±1 µm, respectively. The grain sizes measured from the unimodal samples are in the same range with the ultrafine and coarse grain regions in the bimodal one. True curves of the samples considering the domain between yield stress (YS) and ultimate tensile stress (UTS) have been plotted in Fig. 2(a). The strength dominance is quite clear for ultrafine grain materials while the coarse grains are showing the curve of highest ductility, Fig. 2(a). Also the same figure evidences the rate and ability of strain hardening behaviour as plotted by dashed lines. The ultrafine grained sample showed the maximum values of strain hardening rate compared to others, Fig. 2(a).

In order to assess the participation of each phase in deformation qualitatively, the yield level approach of Polak has been followed [26]. This method is based on the probability density function of the internal critical stresses $f(\sigma_{ic})$ which was deformed to the strain ε for a polycrystal loaded axially with a stress value σ. This critical internal stress followed a distribution in the deformation volumes loaded parallel and can be expressed with the help of a Probability distribution function $f(\sigma_{ic})$ mentioned in Appendix 1.

For further analysis on the nature of the deformation which is supposed to be dominant by ferritic matrix, second derivative of the tensile curves ($d^2\sigma/d\varepsilon^2$) have been done. The ratio of the peak width values have been taken from the second derivative curve for all the three investigated samples, depicted in Fig. 2(b). In Fig. 2(b), sharp yield levels can be seen in the ultrafine grained samples. So, there is a low effective stress as the centroid of this almost binormal distribution is situated at lowest true strain value. This does also mean that due to availability of tiny micro sized stressed domains the internal stress distribution starts from the lowest value. Peaks are also symmetric in nature showing that equal stress levels co existing in the material. Now, coming to the bimodal, the peaks are not so distinct like ultrafine grained steel (Fig. 2(b). The peaks are rather diffused but still one assymetric distribution can be figured out. There are two much coherent type of domains exist that can take strain. The domain having greater peak width is situated at lower true strain value and that of lower peak width and higher intensity situated at higher true strain side. These cosharing of similar yield levels in which the initial lie pretty within the ultrafine levels. It is evident with increasing strain the second yield level emerged and thus taking most of the strain when the true strain crosses 0.06. The course grain nature is also like ultrafine but the strain distribution is much diffused here. The peaks are not sharp but two yield level distributions are clearly evident in case of bimodal microstructure.

A comparative analysis of this peak width has been done using the full width half maxima (FWHM) of this probability distribution function. The results are depicted in Fig. 2(b). It shows that in bimodal microstructure there exist almost equal strain sharing yield levels depicting more or less iso strained type dual phase behaviour and in coarse grained there is a huge difference between the two levels, ultrafine lies in between.

The Schmid factor and the elastic modulus map of the multiple ferrite grains surrounding a void has been shown in Fig. 3. Inset of Fig. 3(a) is showing the enlarged band contrast map of the void portions surrounded by ferrite grains. The colour legends in both the map are indicating the maximum and minimum values of the Schmid factor (0.3 – 0.6) and elastic modulus (90 – 235). For this particular void, the variations in the elastic modulus values of the grains surrounding the voids are more compared to the Schmid factor values, as confirmed by the colour codes in the figures, Fig. 3. Similar observations have been carried out with statistically significant void nucleation sites (10-15 for each sample). All the observations depicted that voids either nucleated on the grain boundaries of two adjacent grains or from the grain boundary triple points. The Schmid factor differences and elastic modulus differences of the void bounded grains have been plotted in Fig. 4.

From the study of local damage initiation phenomena, it is evident that elastic modulus mismatch gives rise to local stress concentration. Quite recently Shakerifard et al [27] quantified the damage initiation in advanced high strength steel and obtained a conclusion that orientation dependence in void initiation may be blurred by presence of different phases in a multiphase microstructure. But here, only a single phase with some prominent granular architecture has been chosen to show the mutual dependence of elastic and plastic incompatibilities. This work has been aimed at formulating a local crystallographic criterion for steels having wide range and distribution of ferritic grains. It is very much understandable that the grain boundaries between incompatible grains give rise to void nucleation. A potential crystallographic description for incompatible grains has been attempted to be formulated and three microstructures (i.e. ultrafine, bimodal and coarse grained) are ranked based on this incompatibility criterion, formulated as the combined contribution of elastic incompatibility and plastic incompatibility. The elastic and plastic incompatibilities have been characterized by local elastic modulus difference and local difference in Schmid factors, respectively. Now, from Fig. 4, it is evident that the sensitivity towards void nucleation in terms of elasto-plastic incompatibility is highest in bimodal microstructure. Slight difference in elastic modulus and Schmid factor gives rise to void initiation in this. From the perspective of void initiation, bimodal structure is highly unstable. In ultrafine grain microstructure, it requires grains with higher and moderate difference in elastic modulus to initiate voids compared to the bimodal one. This microstructure shows a lower degree of instability. Coarse grained microstructure shows lowest degree of instability as it require grains with higher elastic modulus and Schmid factor mismatch to create voids. A probable reasoning can be given for this behaviour based on the processing schedules for the formation of these structures, Fig. 1. Earlier studies showed that the bimodal microstructure is having coarse grains with a dominance of alpha fibre texture *(<110//RD>)* whereas the fine grains embedded in it shows texture randomization [16]. The ultrafine grain sample possessed random texture [20] while cold rolling and annealing texture is expected in coarse grain structure. Considering all the investigated samples, texture randomization occurred because of the retransformation of austenite grains to ultrafine ferrite. The cold / warm rolled ferrite grains got recrystallized followed by growth during the processing schedule (Fig. 1a), leading to the formation of coarse grains. From Fig 4, it is evident that the bimodal grain structures are most sensitive for void formation followed by ultrafine and coarse grain materials. The following factors can be considered in order to address the phenomena; (i) grain boundary misorientation adjacent to void (ii) local strain and texture within the grains and (iii) size difference of the grains surrounding void .In the present investigation, all the samples are recrystallized with more than 80 percent high angle grain boundaries. So, the randomization of texture along with grain size should have played a major role for the initiation of voids.

The probability of texture randomization of grains surrounding the void is highest in case of ultrafine structure. In case of bimodal structure, there is also a decent probability of random texture as the coarse grains are heterogeneously distributed along with the ferrite grains in the bimodal matrix. The probability is lowest in case of coarse grain samples as it possesses predominantly alpha and gamma fibre texture. The more the texture randomization within the neighbourhood grains, the less will be the compatibility among those grains. In this proposed concept, the ultrafine structure should have showed the high sensitivity of the void formation, but the size difference of the grains surrounding the voids played a major role here. Combining both, i.e. grain size and neighbouring grain orientation, the bimodal grain structure showed the highest sensitivity towards void nucleation (Fig. 4). Under uniaxial tension, randomly textured grain aggregates with different sizes will have different dislocation distribution and pileup stresses along with different alignments from the loading axis. The combined effect of two mechanisms probably boosts the bimodal structure to nucleate voids very quickly compared to others. The ultrafine grains are having lowest elongation to failure and the highest UTS. Therefore, the available strain accumulation time to initiate void is lower here and randomly oriented grains enhances the modulus difference or Schmid factor difference requirement. Coarse grain shows maximum stability in this aspect as it is having very little amount of random texture and largest grain sizes among all. This will accommodate more dislocations within it during straining, delaying the void nucleation. Similar type of textures in the neighbouring grains enhances the compatibility of the coarse grain structure.

In the present case, the grains qualifying for void initiation should have an elastic modulus difference 60-80 MPa or if it is lower than 60 MPa it should have a Schmid factor difference of more than 0.1. Therefore, either higher elastic or higher plastic incompatibility is favoured for void nucleation. The aforementioned grain incompatibility has a lower bound also below which no void initiation has been identified. This combinatorial approach of treating elastic and plastic modulus represents a complex dependence of local crystallographic parameters present in the grains. The current representation also clarifies the fact that crystallographic constraints constitute some role behind damage initiation; it is not purely random as classical continuum mechanics describes.

**Acknowledgment**: The authors would like to acknowledge Research & Development, TATA Steel, Jamshedpur for supplying the materials and support from Central Research Facility-IIT Khargapur to carry out experiments.

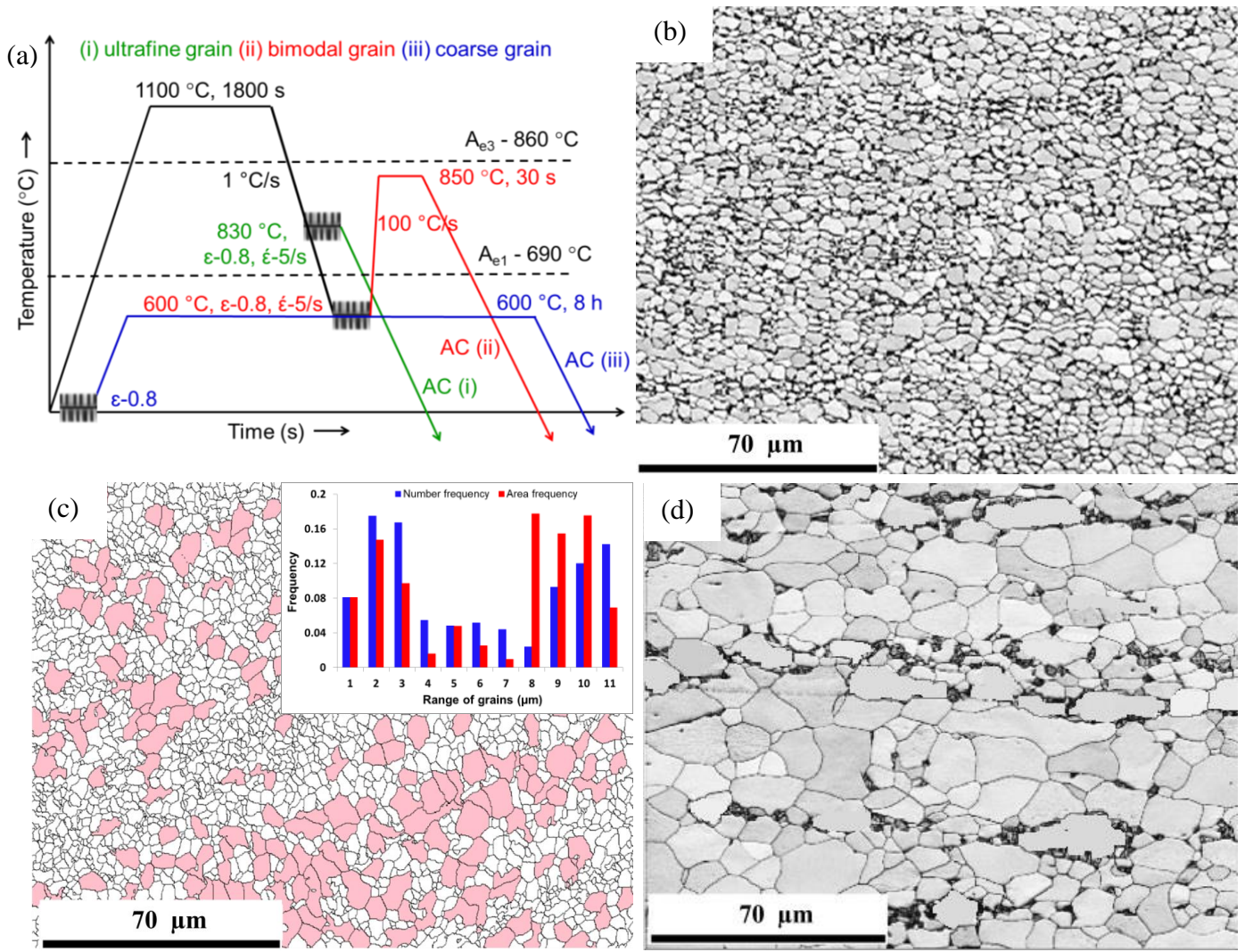

Fig: 1: (a) Schematic diagrams of the heat treatment schedules for the development of different ferrite grain sizes. Band contrast images of ferrite-carbide structures with (b) ultrafine, (c) bimodal and (d) coarse ferrite grain sizes. The coarse grain regions are highlighted separately in the bimodal structure in (c). Number and area frequency evidenced the bimodal grain size distributions in the inset of (c). [AC-air cooling, FC-furnace cooling, ε- amount of strain, έ- strain rate].

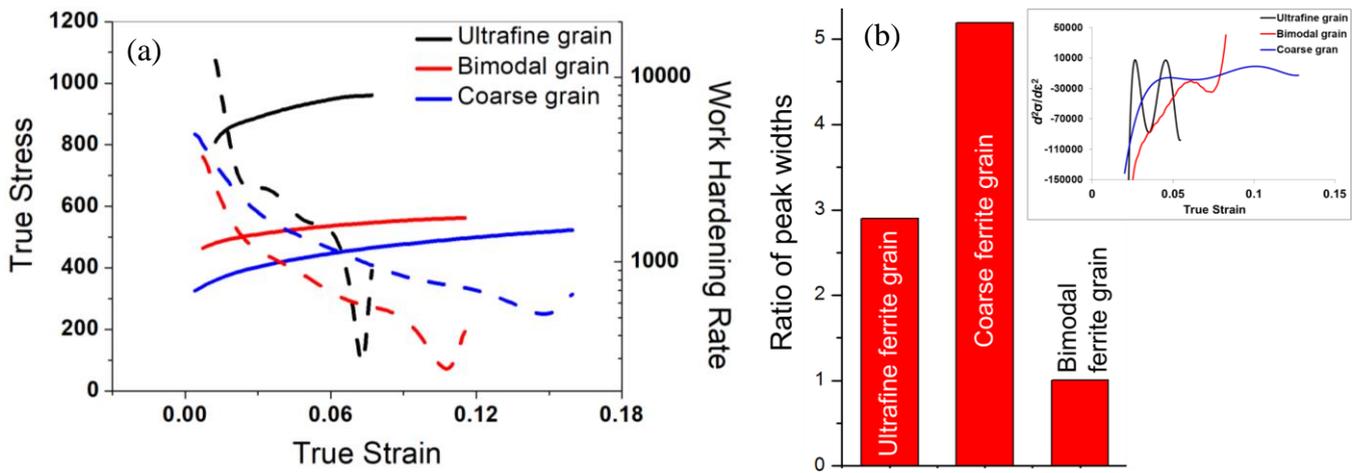

Fig. 2: (a) True stress (solid lines) and work hardening rate (dashed lines) of different grain structures have been plotted against true strain, (b) ratio of peak widths (1$^{st}$ peak : 2$^{nd}$ peak taken from $d^2\sigma/d\varepsilon^2$ vs. $\varepsilon$ plot) for different ferrite structures. The plot of second derivative of the samples are shown n the inset of 2b.

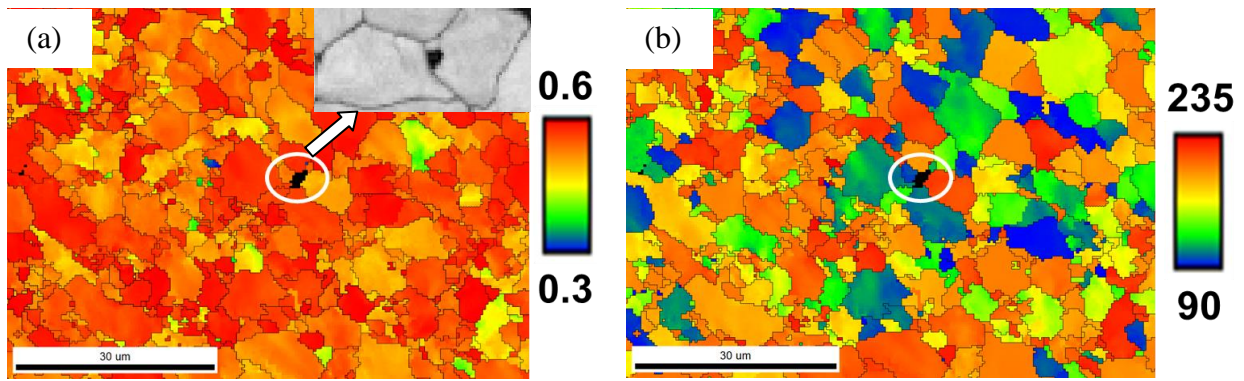

Fig. 3: (a) Schmid factor and (b) elastic modulus map of the respective grains around a void depicted by the black spot in the figures. Inset in (a) showing the enlarged view of the void zone. Attached colour legends are showing the minimum and maximum value of schmid factor and elastic modulus.

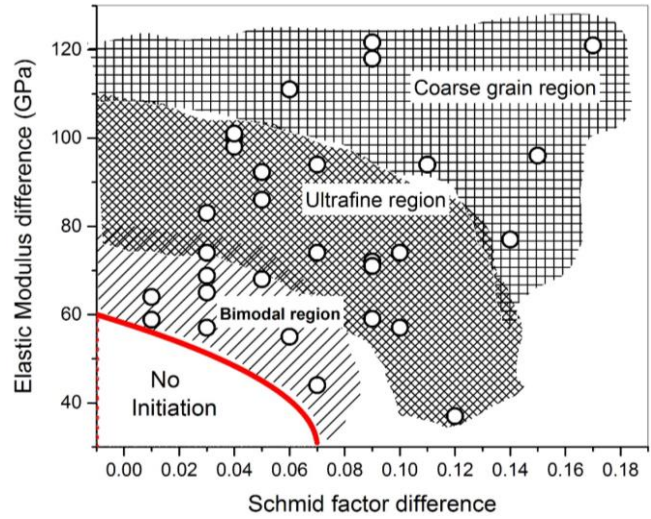

Fig. 4: Plot between elastic modulus difference and Schmid factor difference showing the void initiation regions for different grain structures along with the no initiation zone.

Table 1. Chemical composition (wt. %) of the investigated sample.

| C | Mn | Si | P | S | Nb | Ti | V | Al | N |
|---|---|---|---|---|---|---|---|---|---|
| 0.1 | 1.42 | 0.33 | 0.01 | 0.003 | 0.05 | 0.01 | 0.05 | 0.035 | 0.007 |

## Appendix 1:

An elegant way to understand this continuous distribution of yield is through use of statistical techniques such as probability density function (PDF). A number of previous investigators have used PDF method to explain the local alterations in the yield levels. The method basically describes the distribution of yield levels ($\sigma_{ic}$) of the elements (assuming the material is composed of several parallel elements, each having a different yield level $\sigma_{ic}$. Further details of continuous yield probability can be described below by a probability density function (PDF), $f(\sigma_{ic})$ which upon normalization would yield

$$\int_0^\infty f(\sigma_{ic})d\sigma_{ic} = 1$$

Thus $\int_0^\infty f(\sigma_{ic})d\sigma_{ic}$ represents the area fraction of elements with yield stresses in the interval between $\sigma_{ic}$ and $\sigma_{ic} + d\sigma_{ic}$

Total stress experienced by an axially loaded polycrystalline aggregate is

$$\sigma_{i_{total|}} = \sigma_{eff} + \sigma_{ic}$$

According to the statistical approach, the macroscopic stress σ can be expressed as

$$\sigma = \int_0^{\varepsilon E} \sigma_{ic} f(\sigma_{ic})d\sigma_{ic} + \varepsilon E \int_{\varepsilon E}^\infty f(\sigma_{ic})d\sigma_{ic}$$

Where $f(\sigma_{ic})$ is the probability distribution of the internal stresses in the polycrystalline aggregate showing the Elastic stiffness E.

Now, solving that Integral equation the probability distribution function can be derived, it takes a form like

$$f(\varepsilon E - \sigma_{eff}) = -\frac{1}{E^2}\frac{\partial^2 \sigma}{\partial \varepsilon^2}$$

So, the second derivative of any stress strain response should depict the Probability distribution of possible constituents level those are deformed due to straining.